# Ordinary Chondrite Formation from two Components: Implied Connection to Planet Mercury

J. Marvin Herndon

Transdyne Corporation
San Diego, California 92131 USA

May 12, 2004

## Abstract

Major element fractionation among chondrites has been discussed for decades as ratios relative to Si or Mg. Expressing ratios relative to Fe leads to a new relationship admitting the possibility that ordinary chondrite meteorites are derived from two components: one is a relatively undifferentiated, *primitive* component, oxidized like the CI or C1 chondrites; the other is a somewhat differentiated, *planetary* component, with oxidation state like the reduced enstatite chondrites. Such a picture would seem to explain for the ordinary chondrites, their major element compositions, their intermediate states of oxidation, and their ubiquitous deficiencies of refractory siderophile elements. I suggest that the *planetary* component of ordinary chondrite formation consists of planet Mercury's missing complement of elements.

Key words: ordinary chondrite, meteorite formation, Mercury, element fractionation, asteroid, primordial

Communications: mherndon@san.rr.com

## Introduction

Differences in mean densities of the terrestrial planets have long been interpreted as inferring differences in the major element compositions of the inner planets. As first noted by Urey (1951), the planet Mercury consists mainly of iron; at some point during Mercury's formation, a significant portion of the silicate-forming elements, originally associated with that mass of iron, was lost (Bullen 1952; Urey 1952). Although various mechanisms have been proposed to account for Mercury's great density (Chapman 1988), the ultimate fate of Mercury's complement of lost elements has not, to my knowledge, been addressed.

The constancy in isotopic compositions of most of the elements of the Earth, the Moon, and the meteorites indicates formation from primordial matter of common origin. Primordial elemental composition is yet manifest and determinable to a great extent in the photosphere of the Sun. The less volatile rock-forming elements, present in the outer regions of the Sun, occur in nearly the same relative proportions as in chondritic meteorites. For more than a century, chondrite compositions have been considered relevant to the bulk compositions of the terrestrial planets (Meunier 1871; Daly 1943; Herndon 1980, 1993). But chondrites differ from one another in their respective proportions of major elements (Wiik 1969; Jarosewich 1990), in their states of oxidation (Urey & Craig 1953; Herndon 1996), mineral assemblages (Mason 1962), and oxygen isotopic compositions (Clayton 1993) and, accordingly, are grouped into three distinct classes: *enstatite*, *carbonaceous* and *ordinary*. Understanding how these three distinct classes originated and how they are related to the compositions of planets are among the most fundamental problems in Solar System science.

The ordinary chondrites comprise 80% of the meteorites that are observed falling to Earth. During the 1970s, the widely cited "equilibrium condensation model" was predicated on the assumption that minerals characteristic of ordinary chondrites formed as condensate from a gas of solar composition (Larimer & Anders 1970). Herndon & Suess (1977) demonstrated from thermodynamic considerations, however, that the oxidized iron content of the silicates of ordinary chondrites was consistent instead with their formation from a gas phase depleted in hydrogen by a factor of about 1000 relative to solar composition. Subsequently, I (Herndon 1978) showed that, if the mineral assemblage characteristic of ordinary chondrites could exist in equilibrium with a gas of solar composition, it was at most only at a single low temperature, if at all, and that, moreover, oxygen depletion, relative to solar matter, was also required, as might be expected from the re-evaporation of condensed matter after separation from solar gases.

The abundances of elements in chondrites are expressed in the literature as ratios, usually relative to silicon ($E_i$/Si) and occasionally relative to magnesium ($E_i$/Mg). Expressing Fe-Mg-Si elemental abundances as ratios relative to iron ($E_i$/Fe), as I have done, leads to a relationship bearing on the origin of ordinary chondrites and admitting the possibility that the each ordinary chondrite consists in the main of a mixture of matter from two distinct and reasonably well-characterized reservoirs. The relationship obtained suggests to me the possibility that Mercury's complement of lost elements may comprise one of the two components from which ordinary chondrites appear to have been derived.

**Major Element Relationships**

Only three major rock-forming elements, iron (Fe), magnesium (Mg) and silicon (Si), together with combined oxygen (O) and sulfur (S), account for at least 95% of the mass of each anhydrous chondrite and, by implication, each of the terrestrial planets. Figure 1 presents published analytical whole-rock Fe-Mg-Si data for 206 chondrites, expressed as molar (atom) ratios with respect to iron (Mg/Fe vs. Si/Fe). Major element data for chondrites appear in Fig. 1 to scatter about three distinct, least squares fit, straight lines when plotted as three separate data sets based upon major chondrite classes. The three well-defined straight lines, indicated by dashes in Fig. 1, arise as a result of expressing ratios relative to Fe; chondrite data expressed as ratios relative to Si or to Mg, in the usual manner, will *not* yield three well-defined straight lines, as shown in Table 1.

The existence of three well-defined straight lines for the three sets of chondrite data, shown in Fig. 1, is evident visually (note the expanded detail of the inset) and is demonstrated statistically both by the squares of the correlation coefficients and by the standard error of the least squares regression lines. The standard errors of estimate, calculated for each of the least squares fit straight lines, are shown in Fig. 1 only in the two regions of intersection of the lines. There, intersections of the standard errors of the respective least squares regression lines inscribe parallelograms, indicated in the figure by solid lines, and are indicative of relatively modest variances, thus well-defined straight lines. Figure 1 shows that the major rock-forming elements (Fe-Mg-Si) of individual chondrites within a particular class of chondrites are related in a relatively simple way, as indicated by the fact that data points for members of respective classes scatter about well-defined straight lines. For ordinary chondrite formation, two interpretations are possible: (1) the traditional one-component system; and, (2) the new two-component system I present here.

(1) *One-Component System*: For decades, without benefit of the well-defined straight line relationship shown in Fig. 1, differences in major element compositions of ordinary chondrites have been ascribed to "metal-silicate fractionation" (Larimer & Anders 1970; Ahrens 1964; Kallemeyn, Rubin, Wang & Wasson 1981). In that view, the points that lie along the ordinary chondrite line of Fig. 1 may be considered the result of the simple addition or subtraction of iron metal from its parent material of unspecified origin.

(2) *Two-Component System*: The well-defined straight line relationship shown in Fig. 1 admits a new, different interpretation of ordinary chondrite formation as the result of mixing from two distinct reservoirs. Significantly, in Fig. 1 the ordinary chondrite line intersects both the enstatite chondrite line and the carbonaceous chondrite line in the manner shown. As a consequence, the data for the three major elements (Fe, Mg, and Si) of each ordinary chondrite can be interpreted as lying along a mixing line and consisting of a mixture of two components. One component is defined by the intersection of the ordinary chondrite line and the carbonaceous chondrite line, designated A in Fig. 1 and called *primitive*; the other component is defined by the intersection of the ordinary chondrite line and the enstatite chondrite line, designated B in the same figure and called *planetary*. The elements of the *primitive* source were not previously separated from one another appreciably, like the Orgueil chondrite, whereas the *planetary* source had previously suffered loss of iron metal from a different *primitive*-like parent matter, presumably during proto-planetary core-formation. An important distinction in the new two-component system is that the loss of iron metal, occurs, not in the case of each ordinary chondrite separately, but in the *planetary* reservoir component prior to and perhaps in a different region of space from the location of ordinary chondrite parent formation. In other words, the differences in compositions of individual ordinary chondrites result primarily as a consequence of being formed from different proportions of their two parent components.

The Mg/Fe and Si/Fe ratios of the *primitive* component, point A, can be read from the intersections of the straight lines in Fig. 1 or by simultaneous solution of the equations for the carbonaceous and ordinary chondrite straight lines with slopes and y-intercepts determined from the least squares fit:

*primitive* [Mg/Fe, Si/Fe] = [0.9200, 0.9622]. (1)

Similarly, the Mg/Fe and Si/Fe ratios of the *planetary* component, point B, can be read from line intersections in Fig. 1 or by simultaneous solution of the equations for the enstatite and ordinary chondrite straight lines with their respective slopes and y-intercepts:

*planetary* [Mg/Fe, Si/Fe] = [3.0956, 3.4041]. (2)

Any representative ordinary chondrite can be represented as a mixture of the *primitive* and the *planetary* components. Using (1) and (2) and defining the *primitive* fraction as K and the *planetary* fraction as (1-K), the molar Mg/Fe ratio of any ordinary chondrite is

[Mg/Fe] = 0.9200 K + 3.0956 (1 - K). (3)

Similarly,

[Si/Fe] = 0.9622 K + 3.4041 (1 - K). (4)

The value of K calculated for any particular ordinary chondrite from the ratio [Mg/Fe] differs only by a small amount from the corresponding value of K calculated using the ratio [Si/Fe]. With the intent of smoothing analytical fluctuation, an average of the two values was assigned as the *primitive* fraction, K, of each ordinary chondrite; the *planetary* fraction being (1 - K).

The above approach is justified for the special case of Fig. 1 because Fe, Mg, and Si (together with combined O and S) account for at least 95% of the mass of each chondrite and are fully condensable over a wide variety of conditions. Minor and trace element compositions of the *primitive* and *planetary* components cannot be obtained in the same manner, but estimates can be derived from correlations, provided those elements are fully condensable and are not readily exchanged with an ambient gas phase.

Ordinary chondrite analytical data for many minor and trace elements appear to correlate or to anti-correlate with the planetary fraction as shown in the examples of Fig. 2. Estimates of the ratio $E_i$/Fe of the *primitive* and *planetary* components, obtained from the linear least squares regression line of the ordinary chondrite data as in the examples of Fig. 2 are set forth in Tables 2 and 3.

## Discussion and Conclusions

The *planetary* component, point B in Fig. 1, relative to iron, is enriched approximately three fold in magnesium and silicon compared to the *primitive* component, point A. Such enrichment in silicate-forming elements suggests that the *planetary* component is derived from the outer regions of a partially differentiated proto-planet. For the *planetary* component, one might anticipate similar enrichment in other silicate-forming elements with corresponding deficiencies for iron and for the siderophile elements that dissolve in iron metal, just as shown in Tables 2 and 3.

The approximately seven-fold greater depletion within the *planetary* component of refractory siderophile elements (Ir and Os) than other, more volatile, siderophile elements (Ni, Co and Au), indicates, in at least one instance, that planetary-scale differentiation and/or accretion progressed in a heterogeneous manner. The idea of heterogeneous proto-planetary differentiation and/or accretion is not new. For example, Eucken (1944) suggested core-formation in the terrestrial planets as a consequence of successive condensation on the basis of relative volatility from a hot gaseous proto-planet, with iron metal raining out at the center; Turekian and Clark (1969) reiterated the idea of heterogeneous accretion in the context of a lower pressure, lower temperature model.

There are reasons to associate the highly reduced matter of enstatite chondrites with the inner regions of the Solar System: (1) The regolith of Mercury appears from reflectance spectrophotometric investigations (Vilas 1985) to be virtually devoid of FeO, like the silicates of the enstatite chondrites (and unlike the silicates of other types of chondrites); (2) E-type asteroids (on the basis of reflectance spectra, polarization, and albedo), the presumed source of enstatite meteorites, are, radially from the Sun, the inner most of the asteroids (Zellner, Leake, Morrison & Williams 1977); (3) Only the enstatite chondrites and related enstatite achondrites have oxygen isotopic compositions indistinguishable from those of the Earth and the Moon (Clayton 1993); and, (4) Fundamental mass ratios of major parts of the Earth (geophysically determined) are virtually identical to corresponding (mineralogically determined) parts of certain enstatite chondrites, especially the Abee enstatite chondrite (Herndon 1980, 1993, 1996).

The high bulk density of planet Mercury indicates that much of the silicate matter for the upper portion of Mercury's mantle was lost at some previous time (Urey 1951, 1952; Bullen 1952). I suggest that some matter from the proto-planet of Mercury became the *planetary* component of the ordinary chondrites, presumably separated at the time of Mercury's core formation through dynamic instability and/or expulsion during the Sun's initially violent ignition and approach toward thermonuclear equilibrium.

The *planetary* component of the ordinary chondrites lies along the differentiated portion of the enstatite chondrite line in Fig. 1 and is therefore expected to be highly reduced, like Mercury (Vilas 1985). Moreover, despite great uncertainties, the relative masses involved are all consistent with the possibility that Mercury's complement of lost elements became the *planetary* component of ordinary chondrite formation, re-evaporated together with a more oxidized component of primitive matter, ending up mainly in the region of the asteroid belt, the presumed source region for the ordinary chondrites. Such a picture would seem to explain for the ordinary chondrites, their major element compositions, their intermediate states of oxidation, and their ubiquitous deficiencies of refractory siderophile elements.

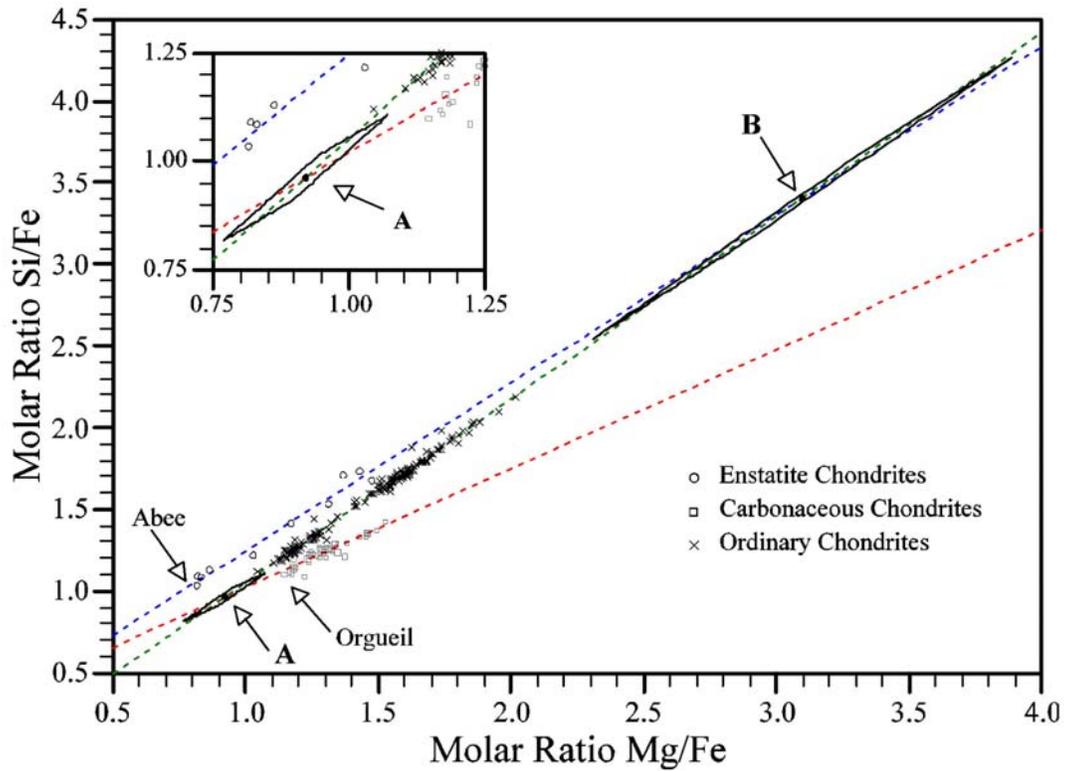

**Figure 1** Molar (atom) ratios of Mg/Fe and Si/Fe from analytical data on 10 enstatite chondrites, 39 carbonaceous chondrites, and 157 ordinary chondrites. Data from Baedecker & Wasson (1975), Jarosewich (1990), Wiik (1969). Members of each chondrite class data set scatter about a unique, linear regression line indicated by dashes. The locations of the volatile-rich Orgueil carbonaceous chondrite and the volatile-rich Abee enstatite chondrite are indicated. Line intersections A and B are designated, respectively, *primitive* and *planetary* components. Error estimates of points A and B are indicated by solid-line parallelograms formed from the intersections of the standard errors of the respective linear regression lines. Inset shows in expanded detail the standard error parallelogram of point A.

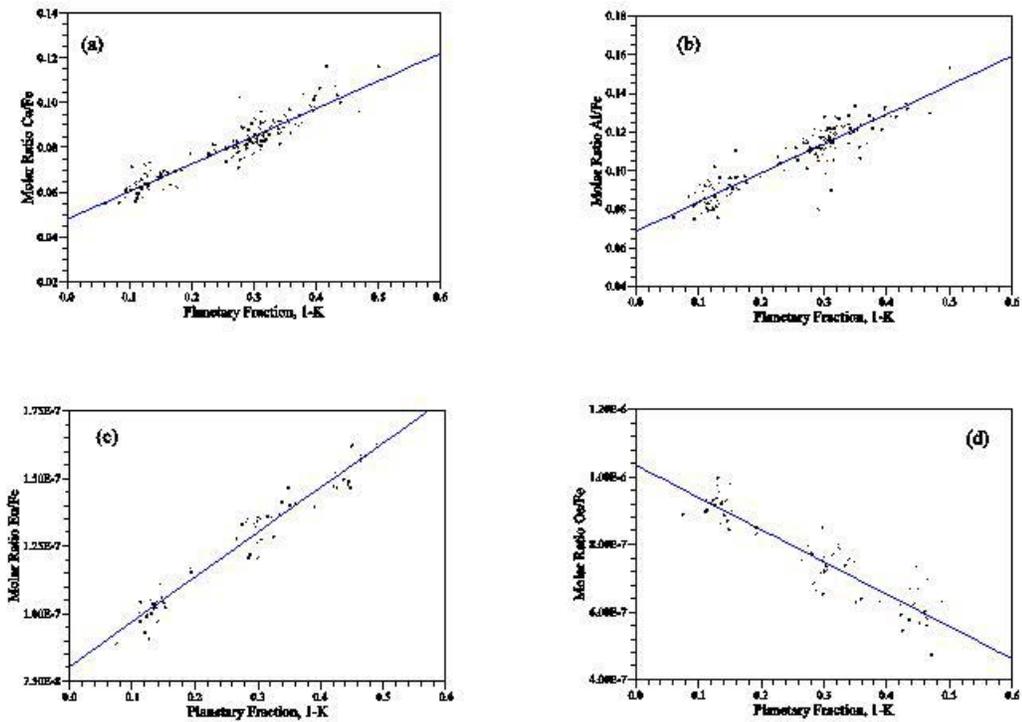

**Figure 2** Examples of ordinary chondrite whole-rock molar (atom) ratios of elements relative to iron as a function of the *planetary* fraction. The square of the correlation coefficient for each of the linear regression lines are as follows: (a) Ca, 0.886; (b) Al, 0.845; (c) Eu, 0.954; (d) Os, 0.846. From analytical data of (Ca and Al) Jarosewich (1990) and (Eu and Os) Kallemeyn, Rubin, Wang & Wasson (1981).

**Table 1** Square of the correlation coefficient of each of the least squares fit lines for the data of Fig. 1 as ratios relative to iron, to silicon, and to magnesium. Note that three well-defined lines occur only in the case of ratios with respect to iron.

|  | Coordinates (Mg/Fe, Si/Fe) Fig. 1 | Coordinates (Mg/Si, Fe/Si) | Coordinates (Si/Mg, Fe/Mg) |
|---|---|---|---|
| **Enstatite Chondrite Line** | 0.973 | 0.505 | 0.663 |
| **Ordinary Chondrite Line** | 0.990 | 0.264 | 0.175 |
| **Carbonaceous Chondrite Line** | 0.814 | 0.005 | 0.250 |

**Table 2** Estimates of elemental (molar) abundance ratios of the *primitive* and *planetary* components of ordinary chondrites for all applicable elements available from the data of Jarosewich (1990). For convenience, the square of the correlation coefficient for the relevant linear least squares regression line is shown in parentheses. For comparison, corresponding elemental abundance ratios are shown for the Orgueil carbonaceous chondrite and the Abee enstatite chondrite.

| Element Ratio | *Planetary* Component B | *Primitive* Component A | Ratio B/A | Orgueil Chondrite | Abee Chondrite |
|---|---|---|---|---|---|
| Mg/Fe (0.997) | 3.0914 | 0.9217 | 3.35 | 1.19 | 0.82 |
| Si/Fe (0.997) | 3.4086 | 0.9609 | 3.55 | 1.08 | 1.11 |
| Ca/Fe (0.886) | 0.1715 | 0.0480 | 3.57 | 0.0802 | 0.0385 |
| Al/Fe (0.845) | 0.2197 | 0.0686 | 3.20 | 0.0956 | 0.0522 |
| Ni/Fe (0.324) | 0.0264 | 0.0648 | 0.41 | 0.0562 | 0.0566 |
| Ti/Fe (0.629) | 0.0074 | 0.0023 | 3.22 | 0.0018 | 0.0014 |
| Mn/Fe (0.807) | 0.0237 | 0.0070 | 3.39 | 0.0140 | 0.0117 |
| Na/Fe (0.630) | 0.1562 | 0.0400 | 3.91 | 0.0673 | 0.0668 |
| Cr/Fe (0.755) | 0.0331 | 0.0113 | 2.93 | 0.0105 | 0.0075 |
| Co/Fe (0.073) | 0.0017 | 0.0029 | 0.59 | 0.0027 | 0.0026 |
| K/Fe (0.452) | 0.0112 | 0.0031 | 3.61 | 0.0045 | 0.0039 |

**Table 3** Estimates of elemental (molar) abundance ratios of the *primitive* and *planetary* components of ordinary chondrites for applicable trace elements from the data of Kallemeyn, Rubin, Wang & Wasson (1981). The *planetary* fraction of each chondrite is calculated from the equation of line intersection, obtained from Fig. 1, using the individual ordinary chondrite molar Mg/Fe ratio. For convenience, the square of the correlation coefficient for the relevant linear least squares regression line is shown in parentheses. For comparison, corresponding elemental abundance ratios are shown for the Orgueil carbonaceous chondrite and the Abee-like Kota-Kota enstatite chondrite from the data of Kallemeyn and Wasson (1981).

| Element Ratio | *Planetary* Component B | *Primitive* Component A | Ratio B/A | Orgueil Chondrite | Abee-like Kota-Kota Chondrite |
|---|---|---|---|---|---|
| Sc/Fe (0.958) | 9.235E-05 | 2.823E-05 | 3.27 | 3.944E-05 | 2.466E-05 |
| V/Fe (0.973) | 7.373E-04 | 2.341E-04 | 3.15 | 3.415E-04 | 2.087E-04 |
| Zn/Fe (0.763) | 3.549E-04 | 1.171E-04 | 3.03 | 1.475E-03 | 8.447E-04 |
| Ga/Fe (0.784) | 3.264E-05 | 1.525E-05 | 2.14 | 4.332E-05 | 4.627E-05 |
| La/Fe (0.936) | 1.142E-06 | 3.490E-07 | 3.27 | 5.053E-07 | 3.499E-07 |
| Sm/Fe (0.941) | 6.442E-07 | 2.058E-07 | 3.13 | 2.780E-07 | 1.678E-07 |
| Eu/Fe (0.954) | 2.459E-07 | 8.055E-08 | 3.05 | 1.165E-07 | 6.806E-08 |
| Yb/Fe (0.955) | 6.506E-07 | 1.891E-07 | 3.44 | 2.945E-07 | 1.793E-07 |
| Lu/Fe (0.958) | 9.219E-08 | 2.954E-08 | 3.12 | 4.186E-08 | 2.483E-08 |
| Au/Fe (0.733) | 1.212E-07 | 2.443E-07 | 0.50 | 2.386E-07 | 3.182E-07 |
| Os/Fe (0.846) | 8.090E-08 | 1.035E-06 | 0.078 | 7.862E-07 | 6.851E-07 |
| Ir/Fe (0.838) | 6.567E-08 | 9.575E-07 | 0.069 | 7.240E-07 | 6.058E-07 |